\documentstyle[12pt]{article}
\begin{document}
\title{Observables of $D=4$ Euclidean Supergravity and Dirac Eigenvalues}
\author{M. A. De Andrade and I. V. Vancea}


\maketitle

The observables of supergravity are defined to be the phase space
gauge invariant objects of the theory. Dirac eigenvalues are
obsevables of Euclidean gravity on a compact manifold [1,2,3]. We
resume the basic results for supergravity [4,5], [6,7].

The general setting which we consider is of minimal supergravity in 
four dimensions on a compact spin manifold without boundary endowed with 
an Euclidean metric $g_{\mu \nu}(x) = e_{\mu}^{a}(x) e_{\nu a}(x)$ where
the indices of the tetrad $e^{a}_{\mu}(x)$ are spacetime indices 
$\mu =1,\cdots ,4 $ and internal Euclidean indices $a=1,\cdots ,4$, respectively.
They are raised and lowered by the Euclidean metric $\delta _{ab}$. 
The gravitino is represented by a Euclidean spin-vector field
$\psi _{\mu}(x)$ and it should be defined
by a modified Majorana condition, since the group $SO(4)$ does not
admit Majorana spinors. A standard condition is $\bar{\psi} = \psi^{T}C$
[8] (see also [9,10]).

By definition, the phase space of the theory is the space of the solutions of the
equations of motion, modulo the gauge transformations  
which are: diffeomorphisms in four dimensions,
local $SO(4)$ rotations and the local $N=1$ supersymmetry. 
The phase space is covariantly defined and its elements 
are all pairs $(e, \psi)$ that are solutions of the equations of motion
modulo gauge transformations. 
Therefore, it is sufficient to consider only on-shell supersymmetry  
and the supersymmetric algebra closes over graviton and gravitino. 
The observables of the theory are the functions on the phase space. 

The self-adjoint Dirac operator and the spin connection are defined as follows
$$
D = i\gamma^a e_{a}^{\mu} (\partial _{\mu} +
\omega _{\mu bc}(e,\psi ) \gamma^b \gamma^c )~~~,~~~
\omega _{\mu ab}(e,\psi ) =  \stackrel{\circ}{\omega _{\mu ab}}(e) + K_{\mu ab}(\psi )   
\eqno{(1)}
$$
where $  \stackrel{\circ}{\omega _{\mu ab}}(e) $ is the usual spin-connection of 
gravity. The following relations hold
$$
\stackrel{\circ }{\omega _{\mu ab} } (e)  =   \frac{1}{2} e_{a}^{\mu}(\partial _{\mu}
e_{b\nu}-
\partial _{\nu}e_{b\mu}) +\frac{1}{2}e_{a}^{\rho}e_{b}^{\sigma}\partial _{\sigma}e_{\rho c}e^{c}_{\mu} -
(a\leftrightarrow b) 
$$
$$
K_{\mu ab}(\psi ) = \frac{i}{4}(\bar{\psi _{\mu}}\gamma _{a}\psi _{b} - 
\bar{\psi _{\mu}}\gamma _{b}\psi _{a} + \bar{\psi _{b}}\gamma _{\mu}\psi _{a}). 
\eqno{(2)}
$$
The chosen topology of spacetime manifold guarantees that
the spectrum of the Dirac operator is discrete
$$
D \chi ^n = \lambda^n \chi ^n  .
\eqno{(3)}
$$
$\lambda^n$'s define a discrete family on the space of 
all gravitons and gravitinos and should be gauge invariant objects. A general 
argument is that the eigenvalues of a general covariant operator should be 
also invariant. By checking up the invariance of $\lambda^n$'s
explicitly one obtains a set of equations that should be
satisfied by the pairs $(e, \psi)$. 

The transformation of the fields under gauge transformations are
given by the following relations
$$
\delta e^{a}_{\mu} = \xi^{\nu} \partial _{\nu} e^{a}_{\mu}~~~,
~~~\delta \psi _{\mu} = \xi^{\nu} \partial _{\nu} \psi _{\mu}, 
\eqno{(4)}
$$
where $\xi = \xi ^{\mu} \partial _{\mu}$ is an infinitesimal vector field,
$$
\delta e^{a}_{\mu} = \theta^{ab} e_{b\mu}~~~,~~~ 
\delta \psi _{\mu}^{\alpha} =  \theta^{ab} (\sigma _{ab})^{\alpha}_{\beta} \psi _{\mu}^{\beta}, 
\eqno{(5)}
$$
where $\theta _{ab}=-\theta _{ba}$ parametrize an infinitesimal $SO(4)$ rotation
and $\sigma ^{ab} = i\Sigma^{ab}$, and
$$
\delta e^{a}_{\mu} = \frac{1}{2}\bar{\epsilon}\gamma^{a}\psi _{\mu}~~~,~~~
\delta \psi _{\mu} = {\cal D} _{\mu}\epsilon ,               
\eqno{(6)}
$$
where $\epsilon (x)$ is an infinitesimal Majorana spinor field, i. e. it obeys
the Majorana conjugation given above. If $\lambda^n$'s are invariant under all 
gauge transformations the following relations must hold
$$
{\cal T}^{n\mu }_{a} \partial _{\nu}e^{a}_{\mu} - 
\Gamma^{n \mu }_{a}\partial _{\nu}\psi _{\mu}^{\alpha} =0,
\eqno{(7)}
$$ 
$$
{\cal T}^{n\mu }_{a}e_{b\mu} +\Gamma^{n\mu }\sigma _{ab}\psi _{\mu}=0, 
\eqno{(8)}
$$
$$
{\cal T}^{n\mu }_{a}\bar{\epsilon}\gamma^{a}\psi _{\mu} + \Gamma^{n\mu }{\cal D}_{\mu}\epsilon =0. 
\eqno{(9)}
$$
Here, ${\cal T}^{n\mu }_{a} = T^{n \mu }_{a} +K^{n\mu }_{a}$,  
where $T^{n\mu }_{a}$  
is the ``energy-momentum tensor'' of the spinor $\chi^n$ 
[2],
$K^{n\mu }_{a}= <\chi^n |i\gamma _{a}K^{\mu }_{bc}(\psi ) \sigma^{bc}|\chi^n >$,
and
$$
\Gamma^{n \mu }_{a} = \frac{i}{4}\int \sqrt{e} {\chi^n}^{\ast}\gamma^a e_{a}^{\nu}
[\bar{\psi_{\nu}^{\beta}}(\gamma _{b})_{\alpha \beta }e^{\mu}_{c} -
\bar{\psi_{\nu}^{\beta}}(\gamma _{c})_{\alpha \beta }e^{\mu}_{b} +
\bar{\psi_{b}^{\beta}}(\gamma _{\nu})_{\alpha \beta }e^{\nu}_{c}]\sigma^{bc} \chi^{n}. 
\eqno{(10)}
$$
Eq.(5) is a covariant equation, therefore its variation 
under gauge transformations should cancel. The variation 
of Eq.(5) under diffeomorphisms, $SO(4)$ rotations and local 
supersymmetry and the equations (7), (8) and (9) imply the 
following relations 
$$
\{ [b^{\mu}(\xi ) - c(\lambda \xi )^{\mu} ]\partial _{\mu} + f(\xi ) \} \chi^n = 0, 
\eqno{(11)}
$$
$$
[\theta_{a}^{a}D -g(\theta ) +h(\theta ) ]\chi^n = 0, 
\eqno{(12)}
$$
$$
[j^{\mu}_{a} (\epsilon ) \partial_{\mu} + k_{a} (\epsilon ) +l_{a} ]\chi^n =0. 
\eqno{(13)}
$$
Here, the following notations are used
$$
h(\theta ) =  i (\lambda^n - D) {\bf \theta \sigma}
~~~,~~~
j^{\mu }_{a} (\epsilon )  =  \frac{1}{2}\gamma_{a}\bar{\epsilon}\psi^{\mu} ~~~,
~~~k_{a}(\epsilon ) = 
\frac{1}{2}\gamma_{a}\bar{\epsilon}\psi^{\mu}\omega_{\mu cd}\sigma^{cd}
\eqno{(14)}
$$
$$
b^{\mu}(\xi )  = i \gamma^{a} b_{a}^{\mu}(\xi )~~~,~~~  
b_{a}^{\mu}(\xi ) = \xi^{\nu}\partial_{\nu}e_{a}^{\mu} -e_{a}^{\nu}\partial_{\nu}\xi^{\mu}
-2e_{a}^{\nu}\xi^{\mu}\omega_{\nu bc}\sigma^{bc}
\eqno{(15)}
$$
$$  
c(\lambda ,\xi )^{\mu} = (\lambda^n - D)\xi^{\mu}~~~,~~~   
f(\xi ) = i\gamma^{a}\xi^{\nu}\partial_{\nu}(e_{a}^{\mu}\omega_{\mu bc})\sigma^{bc}, 
\eqno{(16)}
$$
$$
g(\theta ) = [\gamma^c e_{c}^{\mu}([{\bf \theta \sigma},\omega_{\mu ab}] -
\partial_{\mu}{\bf \theta \sigma }M_{ab})]\sigma^{ab},
\eqno{(17)}
$$
$$
l_a =e_{a}^{\mu}[B_{\mu cd} - \frac{1}{2}e_{\mu d}B_{ec}^{e} +
\frac{1}{2}e_{\mu c}B_{ed}^{e}]\sigma^{cd}.
\eqno{(18)}
$$
We interpret the relations (7), (8), (9) 
and (11), (12), (13) as two sets of 
constraints on the phase space of the theory. By construction, the second 
set is determined by the first one. An analysis of the linear dependency 
of these constraints in view of the quantization of the theory was
performed in [4,5,6,7] where some other aspects of the theory were 
discussed. A "spectral formulation" of supergravity is not known at present.

\newpage

{\bf BIBLIOGRAPHY}:\\

[1] G. Landi and C. Rovelli, Phys. Rev. Lett.78(1997)3051, 
gr-qc/9612034;\\ 

[2] G. Landi and C. Rovelli, Mod. Phys. Lett.A13 (1998)479-494,
gr-qc/9708041;\\ 

[3] G. Landi, gr-qc/9906044;\\

[4] I. V. Vancea, Phys. Rev. Lett.79(1997)3121-3124, gr-qc/9707030; 
Err. ibid.80(1998)1355;\\ 

[5] I. V. Vancea, Phys. Rev. D58(1998)045005, gr-qc/9710132;\\

[6] N. Pauna and I. V. Vancea, Mod. Phys. Lett. A13(1998)3091-3098, 
gr-qc/9812009;\\

[7] C. Ciuhu and I. V. Vancea, Int. Journ. Mod. Phys.A15(2000)2093-2104, 
gr-qc/9807011; \\

[8] P. Van Nieuwenhuizen, in {\em "Relativity, Groups and Topology
II"}, Proceedings of the Les Houches Summer School, 1983, edited by R. Stora 
and B. S. DeWitt, Les Houches Summer School Proceedings Vol40 (North-Holland, 
Amsterdam, 1984)\\

[9] J. Kupisch and W. D. Thacker, {\em Fortschr. Phys.}38, 35(1990);\\

[10]G. Esposito, {\em Complex General Relativity}( Kluwer Academic 
Publishers, 1995)\\

\end{document}